\documentclass[colorlinks]{elsart}
\usepackage{amsmath, amsfonts}

\usepackage{graphicx}
\usepackage{subfigure}
\usepackage{overpic}

\usepackage{tikz}
\usetikzlibrary{arrows,shapes}

\definecolor{dgreen}{rgb}{0,.6,0}
\newtheorem{thm1}{Theorem}
\newtheorem{cor1}{Corollary}[thm1]
\newtheorem{defn1}{Definition}
\newtheorem{prop1}{Proposition}

\usepackage[breaklinks=true,a4paper=true,pagebackref=false]{hyperref}
\begin{document}
% Define block styles
\tikzstyle{decision} = [diamond, draw, fill=blue!20,
    text width=4.5em, text badly centered, node distance=3cm, inner sep=0pt]
\tikzstyle{block} = [rectangle, draw, fill=blue!20,
    text width=5em, text centered, rounded corners, minimum height=4em]
\tikzstyle{line} = [draw, -latex'] \tikzstyle{cloud} = [draw,
ellipse,fill=red!20, node distance=3cm,
    minimum height=2em]

\newlength\figwidth
\setlength\figwidth{0.5\columnwidth}
\newlength\imgwidth
\setlength\imgwidth{0.3\columnwidth}

\begin{frontmatter}
\title{Application of Gray codes to the study of the theory of symbolic dynamics of unimodal maps}
\author[Spain]{David Arroyo\corauthref{corr}}
and
\author[Spain2]{Gonzalo Alvarez}

\corauth[corr]{Corresponding author: David Arroyo
(david.arroyo@uam.es).}
\address[Spain]{Grupo de Neurocomputaci\'on Biol\'ogica, Dpto. de Ingenier\'ia Inform\'atica. Escuela Polit\'ecnica Superior. Universidad Aut\'onoma de Madrid, 28049 Madrid, Spain
}
\address[Spain2]{Instituto de Seguridad de la Informaci\'on, Consejo Superior de
Investigaciones Cient\'{\i}ficas, Serrano 144, 28006 Madrid, Spain}

\begin{abstract}
In this paper we provide a closed mathematical formulation of our
previous results in the field of symbolic dynamics of unimodal
maps. This being the case, we discuss the classical theory of
applied symbolic dynamics for unimodal maps and its reinterpretation
using Gray codes. This connection was previously emphasized but no
explicit mathematical proof was provided. The work described in
this paper not only contributes to the integration of the different
interpretations of symbolic dynamics of unimodal maps, it also
points out some inaccuracies that exist in previous works.
\end{abstract}
\end{frontmatter}
\section{Introduction}
A symbolic sequence is a transformation of a sequence of real numbers
into a sequence consisting of a set of symbols. Regarding unimodal
maps, the cardinality of that set is two and it is determined by the turning
point of the iteration function of the map. Accordingly, each symbol
represents the relative position of a real-value with
respect to the turning point. In \cite{metropolis73} it is pointed out
the existence of an inner order of the symbolic sequences, along with the relationship between this order and
the initial condition and the control parameter of the underlying
chaotic system. The considerations and results of \cite{metropolis73} were later improved
and enlarged through different contributions, being the most important
\cite{BMS86} and \cite{wang87}. In \cite{alvarez98} it was remarked
that the order of the symbolic sequences can be interpreted using the
concept of Gray codes. In this novel approach to the problem, the
symbolic sequences are finally converted into a figure which is a real
number between $0$ and $1$ called Gray Ordering Number or simply
GON. Afterwards , \cite{cusick99} drew the bridge between the ideas of
\cite{alvarez98} and the main theory of applied symbolic dynamics as
expressed in \cite{wang87}. Finally, some theorems are offered in
\cite{wu04}, which enlarge the theoretical framework of the GON of
unimodal maps. In \cite{wu04} it is explained that the dynamical
properties of unimodal maps by means of the GON are a translation of
the theoretical framework inherited from \cite{metropolis73}.
Nevertheless, there is no direct and explicit proof of this
equivalence. One of the main applications of the concept of the GON is
the estimation of the control parameter of unimodal maps for
cryptanalysis \cite{alvarez03a,arroyo09,rhouma09,arroyo11}. The precise
definition of the key space of a cryptosystem is a commitment
in cryptography. In the context of chaotic cryptography, it implies
that the control parameters and initial conditions of
the chaotic system must be selected to guarantee chaoticity, and to
avoid the estimation of either control parameters or initial
conditions from partial information about the chaotic orbits
\cite[Rule 5]{Alvarez06a}. In case that this partial information arises
from the symbolic sequences of the chaotic map used for encryption,
we must assess that it is not possible to get an accurate enough
estimation of control parameters and/or initial conditions. Therefore,
a rigorous and concrete theoretical framework is required to quantify
the precision of the procedures for the estimation of the control
parameter and the initial condition of unimodal maps from their
symbolic sequences. This paper presents this
concretion and also shows that some of the theorems in \cite{wu04} are
not totally accurate. In this sense, those theorems are not only
criticized but also rewritten.

This paper is organized as follows. First of all, Sec.
\ref{section:scenario} introduces the class of maps under study and
the main aspects of their symbolic dynamics. Section
\ref{section:initialCondition} remarks the existence of an inner
order for the symbolic sequences of a certain class of unimodal maps
and a relationship between that order and the order of the initial
conditions employed in their generation. In Sec.
\ref{section:grayCodes} the order of the symbolic sequences is
rewritten in terms of Gray codes and the concept of Gray Ordering
Number is introduced. After that, Sec.
\ref{section:controlParameter} introduces a subclass of the class of
considered unimodal maps. This subclass of unimodal maps is defined
in a parametric way, i.e., their dynamics depend on a control
parameter. This dependency is analyzed by means of the GON. This
study will lead to the revision and proof of all theorems in
\cite{wu04}. Finally, Sec. \ref{section:conclusions} summarizes the
main results of the present work.

\section{Scenario}
\label{section:scenario}
The work described in this paper is focused
on a special class of functions. This class is denoted by
$\mathcal{F}$. A function $f$ belonging to the class $\mathcal{F}$
is defined in the interval $I=[a,b]$ for $a<b$ and satisfies:
\begin{enumerate}
    \item $f$ is a continuous function in $I$.
    \item $f(a)=f(b)=a$.
    \item $f(x)$ reaches its maximum value $f_{\max}\leq b$ in the
    sub-interval $[a_m,b_m]\subset I$ so that $a_m\leq b_m$.
    \item $f(f_{\max})< x_c$ and $f(f_{\max}) \geq a$, where $x_c$ is the middle point of the interval $[a_m,b_m]$ , i.e.,
    $x_c=\frac{a_m+b_m}{2}$.
    \item $f(x_c)>x_c$
    \item $f(x)$ is an strictly increasing function in $[a,a_m]$ and
    an strictly decreasing function in $[b_m,b]$.
\end{enumerate}

\begin{figure}[!htbp]
\centering
\includegraphics{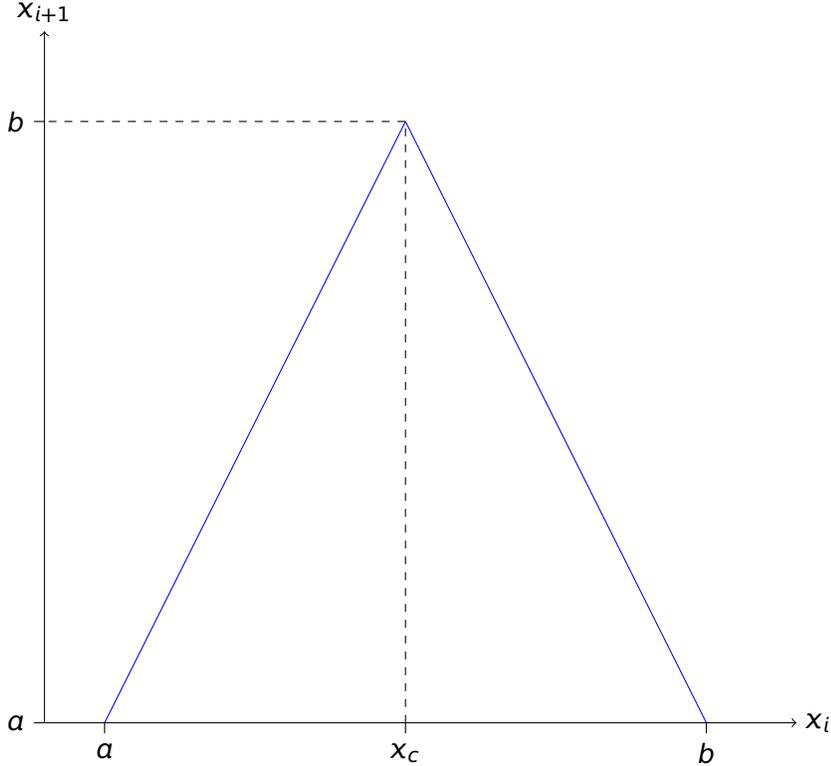}
\caption{Tent map.} 
\label{figure:tentMap}
\end{figure}
Although the work in this paper is focused on the class of functions
$\mathcal{F}$, it is possible to extend it to other class of
functions considering the topological conjugacy of maps \cite[p.
72]{hao:book98}. This other class of functions is named
$\mathcal{F}^*$ and any $f$ included in $\mathcal{F}^*$ has the same
properties as those in $\mathcal{F}$ with the exception of
properties $(3)$ and $(6)$, since if $f$ is in  $\mathcal{F}^*$,
then it possesses a minimum value in $[a_m,b_m]$ and is strictly
decreasing in $[a,a_m]$ and strictly increasing in $[b_m,b]$.

Hereafter, the function $f(x)$ is considered as a way to generate a
sequence of numbers $\left\{x_i \right\}$ from a certain initial
value $x_0$. Each number $x_{i}$ determines the next element of the
sequence trough $x_{i+1}=f(x_i)$. After a transient number of
iterations, all the $x_i$ values are inside the interval
$[x_{\min},x_{\max}]$, where $x_{max}=f(x_c)$ and
$x_{min}=f(x_{\max})$.

The tent map is included in the class $\mathcal{F}$ and is
represented in Fig. \ref{figure:tentMap}. In this case $a_m=b_m=x_c$
and $f_{\max}=f(x_c)=b$. A certain value $x_{i+1}\neq x_c$ can be
derived from two different values of $x_i$, as
Fig.\ref{figure:tentMap} informs. In other words, it is satisfied
that $x_{i+1}=f(x_i^L)=f(x_i^R)$, where $x_i^L\neq x_i^R$,
$x_i^L<x_c$ and $x_i^R>x_c$. This is a common characteristic of all
the functions of the class $\mathcal{F}$. It means that the initial
condition used in the generation of $\left\{x_i\right\}$ using
$f(x)$ can be recovered from the last number of the sequence only if
the relative position of every $x_i$ with respect to $x_c$ is known.
Therefore, the recovering of the initial condition demands recording
those relative positions. This is achieved by transforming
$\left\{x_i\right\}$ into a symbolic sequence or pattern according
to the next criterion:
\begin{eqnarray}
    x_i& \equiv & L \text{ if } x_i\in[a,x_c),\\
    x_i&\equiv & C \text{ if } x_i=x_c,\\
    x_i& \equiv & R \text{ if } x_i\in(x_c,b].
\end{eqnarray}
If $f$ is in $\mathcal{F}^*$ instead of being in $\mathcal{F}$, then
the symbolic sequences are generated in the same but changing all
the $L$'s into $R$'s and viceversa.

Consequently, $\left\{x_i\right\}$ is associated to the symbolic
sequence $P=p_0p_1\ldots$ where $p_i\in\left\{L,R\right\}$. Using
$P$ and the last element of $\left\{x_i\right\}$ one can recover the
initial condition $x_0$.

\section{Relationship between the symbolic sequences and the initial condition used in their generation}
\label{section:initialCondition} 
\begin{figure}[!htbp]
    \centering
    \includegraphics{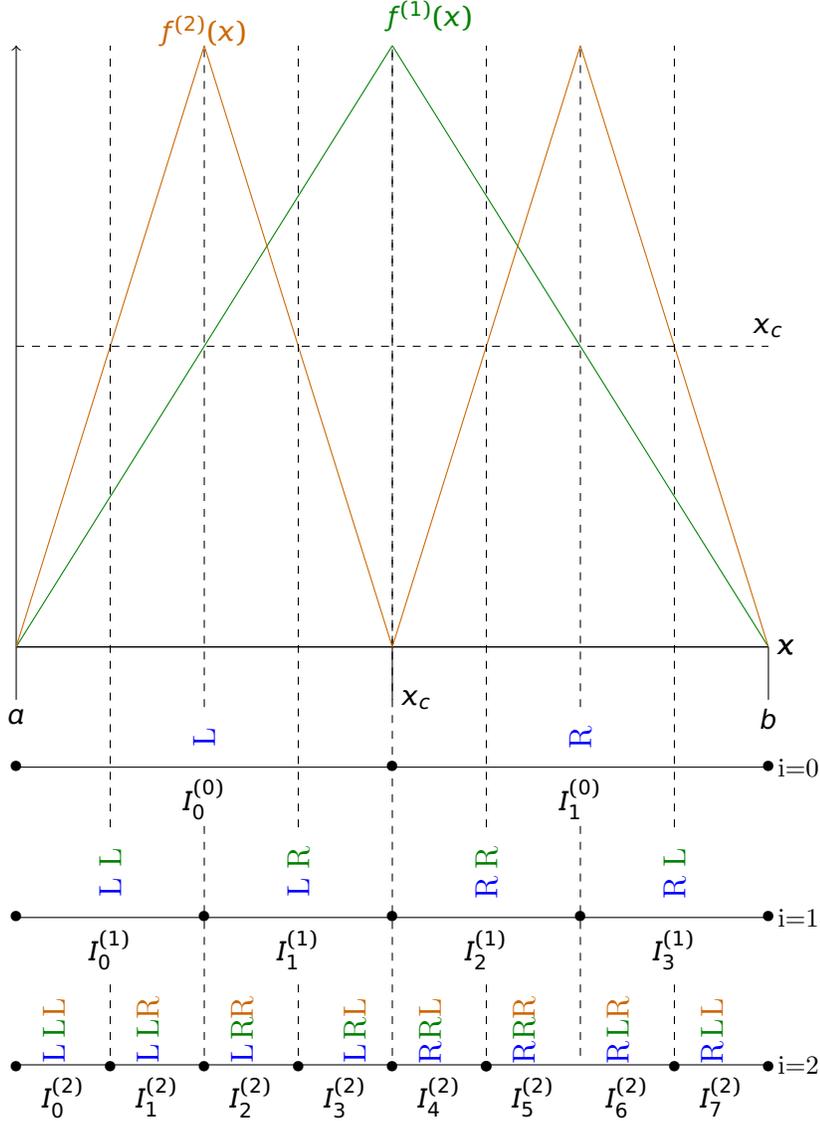}
\caption{Symbolic intervals for different iterations of the tent
map.} \label{figure:tentMapSymbolic}
\end{figure}

Let us call $P_f(x_0)$ to the
symbolic sequence of length $n$ generated from $x_0$ using the
function $f(x)$, which is included in the class $\mathcal{F}$. The
value of the $i-$th symbol of the symbolic sequence $P_f(x_0)$ is
determined by $f^{(i)}(x_0)$, i.e., the $i-$th iteration of $f(x)$
from $x_0$ for $i\in[0,n-1]$. If $p_i$ is the $i-$th symbol of the
symbolic sequence, $p_i$ is equal to $L$ if and only if
$f^{(i)}(x_0)<x_c$. In the same way, $p_i$ is equal to $R$ if and
only if $f^{(i)}(x_0)>x_c$. As a consequence, the definition
interval $I$ is divided into $2^{i+1}$ \emph{symbolic} sub-intervals.
Indeed, if $x_c^{(i,j)}$ is the $j$-th solution of the equation
\begin{equation}
 f^{(i)}(x) = x_c,
 \label{equation:symbolicIntervals}
\end{equation}
the set $\left\{x_c^{(i,j)}\right\}$ for $0\leq j < 2^{i}$ divide
the definition interval into $2^{i+1}$ sub-intervals, where
$x_c^{(0,0)}=x_c$. All the values included in one of these intervals
generate the same symbolic sequence of length $i+1$. In Fig.
\ref{figure:tentMapSymbolic} the symbolic intervals of the tent map
for zero, one and two iterations are depicted. The main result of
the previous proposition is that, for a certain number of
iterations, the different sub-intervals are so that two neighboring
sub-intervals lead to the same symbolic sequence except for one
symbol. On the other hand, for $i\in \left\{0,1,2,\ldots\right\}$
and $j\in [0,2^i-1]$, the set of points $x_c^{(i,j)}$ determine
periodic symbolic sequences of period $i+1$ when they are considered
as initial conditions. If the symbol $C$ is assigned to $x_c$ and
only one period is regarded, the symbolic sequences generated from
$\left\{x_c^{(i,j)}\right\}$ end with a $C$. In this sense, if the
iteration process associated to the generation of a symbolic
sequence stops just when a $C$ is obtained, only the symbolic
sequences derived from the set of initial conditions solution of
Eq.~\eqref{equation:symbolicIntervals} have finite length.

All the previous observations can be formally expressed by the
following definition:

\begin{defn1}
    For a certain function $f(x)$ the symbolic sequence or kneading sequence
    generated from the initial condition $x_0$ is $P_{f}(x_0)$. If exists $i\in \mathbb{N}_0$
    such that $f^{(i)}(x_0)=x_c$, then $P_{f}(x_0)$ is finite length. Otherwise, $P_{f}(x_0)$ is a kneading
    sequence of infinite length. As a consequence, any kneading
    sequence of finite length always ends with a $C$.
\end{defn1}

If $\mathcal{S}$ is the set of all sequences derived from the
iteration of the functions included in $\mathcal{F}$, then it is
possible to derive a complete ordered set
$\left(\mathcal{S},<_{\mathcal{S}}\right)$ where the referred order
is defined according to \cite[p. 309]{BMS86} as follows:
\begin{lem}
    Assuming $L<_{\mathcal{S}}C<_{\mathcal{S}}R$, $S=\left\{s_i\right\}$ and
$T=\left\{t_i\right\}$ $\in \mathcal{S}$, and $j$ is the first index
so that $s_j\neq t_j$, it is said that $S<_{\mathcal{S}}T$ if one of
the next conditions is satisfied:
    \begin{enumerate}
        \item $j=0$ and $s_0<_{\mathcal{S}}t_0$.
        \item $j>0$, $s_0 s_1\ldots s_{j-1}= t_0 t_1\ldots t_{j-1}$
              contains an even number of $R$'s and $s_j<_{\mathcal{S}}t_j$.
        \item $j>0$, $s_0 s_1\ldots s_{j-1}= t_0 t_1\ldots t_{j-1}$
              contains an odd number of $R$'s and $s_j>_{\mathcal{S}}t_j$.
    \end{enumerate}
\label{lemma:order}
\end{lem}

The inner order of $\mathcal{S}$ is directly linked to the order on
$\mathbb{R}$ of the real numbers in $I$ used to generate the
symbolic sequences from any $f$ in $\mathcal{F}$. This is informed
and proved in \cite[Lemma 4.1]{BMS86} and in \cite[Theorem
2]{wang87}. For the sake of clarity, the relationship between the
order of the kneading sequences and the order of the initial
conditions is rewritten as a theorem:
\begin{thm1}
    For $f(x)$ belonging to the class of functions $\mathcal{F}$ and
    $x,y$ included in the interval of definition of $f(x)$ so that $x<y$, it is
    verified that $P_f(x)\leq_{\mathcal{S}} P_f(y)$.
    \label{theorem:order0}
\end{thm1}

\section{Gray codes and symbolic sequences}
\label{section:grayCodes} In the previous section it was remarked
that $f^{(n)}(x)$ can be divided into $2^{n+1}$ intervals such that
all the values included in one of those intervals lead to the same
symbolic sequence of length $n+1$. In this sense, those intervals
were referred as symbolic intervals, since a certain interval can be
named through the symbolic sequence generated from any value inside
it. It was also observed that two contiguous symbolic sequences
differed in just one symbol. Finally, if the first symbol of the
symbolic sequences is discarded, the $2^{n+1}$ symbolic sub-intervals
generated by the $n-$th iteration of the map $f(x)$ are symmetric
with respect to $x=x_c$. In communication theory it is very well
known a family of codes distinguished by the fact that two
successive codes only differ in one bit. This family of codes is the
Gray codes family, which also presents the above cited mirroring
property. Table \ref{table:Graycodes} shows the Gray codes of length
$4$. As a result, it is immediate the translation of the symbolic
sequences of the class of functions $\mathcal{F}$ into binary
sequences just changing the symbol $L$ into $0$ and the symbols $R$
and $C$ into $1$ \cite{alvarez98}. In this sense, the Gray code
associated to a certain pattern $P_f(x)$ is given by the next
definition.

\begin{defn1}
    The Gray code corresponding to $P_f(x)=p_0p_1\cdots p_{j-1}\cdots$ is defined as $G(P_{f}(x))=g_0g_1\cdots
g_{j-1}\cdots$ where
    \begin{equation*}
        g_i=\left\{\begin{array}{cc}1&\text{ if } p_i=R\\ 0 & \text{ if } p_i=L, \end{array}\right.
    \end{equation*}
    for $i \in \mathbb{N}_0$. If $p_j=C$ for any $j$ in $\mathbb{N}_0$, then the Gray code
    associated to $P_f(x)$ is $g_0g_1\cdots g_j$.
\end{defn1}

As Table \ref{table:Graycodes} informs, it is possible to translate
a Gray code into a binary code. The equivalent binary code of a
given Gray code can be easily obtained using the next definition:

\begin{defn1}
    If the Gray code of a certain symbolic sequence $P_f(x)=p_0p_1\cdots
p_{j-1}\cdots$ is given by $G(P_{f}(x))=g_0g_1\cdots
g_{j-1}\cdots$, then the binary code related to $P_f(x)$ is
$U(P_f(x))=u_0u_1\cdots u_{j-1}\cdots$ where
    \begin{equation*}
        u_{i+1}=u_i\oplus g_{i+1},
    \end{equation*}
for $i \in \mathbb{N}_0$ and $u_0=g_0$. If $P_f(x)$ is of length
$j$, i.e., if $p_{j-1}=C$, then the binary coded related to $P_f(x)$
is $U(P_f(x))=u_0u_1\cdots u_{j-1}$  where
    \begin{equation*}
        u_{i+1}=\left\{\begin{array}{ll}u_i \oplus g_{i+1},& \mbox{ for } 0<i<j-1,\\ 1, & \mbox{ for } i=j-1. \end{array}\right.
    \end{equation*}

\label{proposition:gray}
\end{defn1}

Since a binary code can be interpreted as a decimal number just
changing the base, it is possible to associate a number to a
symbolic sequence. However, the canonical base changing makes the
first symbol modify its weight as the length of the symbolic
sequence increases. In order to avoid the changing of the symbol
weights as the length of the symbolic sequences increases, the Gray
code associated to a symbolic sequence is interpreted as a decimal
number with integer part equal to zero. The next definition
introduces how to carry out the transformation of a symbolic
sequence into a real number between $0$ and $1$.

\begin{table}[htbp]
\centering
\begin{tabular}{|c|c|c|}
 \hline
 \textbf{Rank} &  \textbf{Binary code} &  \textbf{Gray code} \\
 \hline
 0 &   0000 &  0000 \\
 \hline
 1 &   0001 &  0001 \\
 \hline
 2 &   0010 &  0011 \\
 \hline
 3 &   0011 &  0010 \\
 \hline
 4 &   0100 &  0110 \\
 \hline
 5 &   0101 &  0111 \\
 \hline
 6 &   0110 &  0101 \\
 \hline
 7 &   0111 &  0100 \\
 \hline
 8 &   1000 &  1100 \\
 \hline
 9 &   1001 &  1101 \\
 \hline
 10 &  1010 &  1111 \\
 \hline
 11 &  1011 &  1110 \\
 \hline
 12 &  1100 &  1010 \\
 \hline
 13 &  1101 &  1011 \\
 \hline
 14 &  1110 &  1001 \\
 \hline
 15 &  1111 &  1000 \\
 \hline
\end{tabular}
\caption{Correspondence between Gray codes and binary codes for four
bits.} \label{table:Graycodes}
\end{table}

\begin{defn1}
    Let $G(P)=g_0g_1\cdots g_{n-1}$ be a set of bits representing a
Gray code of length $n$. Let $U(P)=u_0u_1\cdots u_{n-1}$ be the
binary code corresponding to $G(P)$. The \textbf{Gray Ordering
Number} or \textbf{GON} of $P$ is defined as the real number given
by
\begin{equation*}
    GON(P)=2^{-1}\cdot u_0 + 2^{-2}\cdot u_1+\cdots +2^{-n}\cdot
u_{n-1}.
\end{equation*}
\end{defn1}

The definition of the GON also implies the definition of an order
$<_{GON}$ upon the set of symbolic sequences $\mathcal{S}$. In other
words, according to the definition of the GON, it is possible to
build the complete ordered set $(\mathcal{S},<_{GON})$. This ordered
set is equivalent to $(\mathcal{S}, <_{\mathcal{S}})$, i.e., the
order defined using the GON is equivalent to the order
$<_{\mathcal{S}}$.

\begin{prop1}
 The orders $<_{\mathcal{S}}$ and $<_{GON}$ are equivalent on
 $\mathcal{S}$.
 \label{corollary:equivalenceWangGray}
\end{prop1}

\begin{pf}
Let $P=p_0p_1\ldots p_{j-1}p_j \ldots$ be a certain symbolic
sequence which can be of finite or infinite length. If
$U(P)=u_0u_1\ldots u_{j-1}u_{j}\ldots$ is the binary code linked to
the kneading sequence $P$, $u_{j}$ is equal to $1$ if one of the
next situations occurs:
    \begin{enumerate}
        \item $p_j=R$ and $p_0p_1\ldots p_{j-1}$ contains an even
        number of $R$'s.
        \item $p_j=L$ and $p_0p_1\ldots p_{j-1}$ contains an odd
        number of $R$'s.
    \end{enumerate}
Let $Q=q_0q_1\ldots q_{k-1} q_{k}\ldots $ be another kneading
sequence of finite or infinite length. Let $U(Q)=t_0t_1\ldots
t_{k-1}t_{k}\ldots$ be its associated binary code. According to
Theorem \ref{theorem:order0}, if the first different symbol between
$P$ and $Q$ is the $i-$th one, then $P<_{\mathcal{S}}Q$ if and only
if one of the next cases happens:
    \begin{enumerate}
        \item $p_i=R$, $q_i=L$ and $p_0p_1\ldots p_{i-1}$ contains an
        odd number of $R$'s. As a consequence, it is verified that $u_i=0$  and
        $t_i=1$, which implies that $GON(P) < GON(Q)$, i.e.,
        $P<_{GON} Q$.
        \item $p_i=R$, $Q$ of length $i$ and $p_0p_1\ldots p_{i-1}$ contains an
        odd number of $R$'s. Since $Q$ is finite-length, its final
        symbol is $C$. Therefore, $t_i=1$ and $u_i=0$ implying that
        $GON(P)< GON(Q)$, i.e., $P<_{GON} Q$.
        \item $p_i=L$, $q_i=R$ and $p_0p_1\ldots p_{i-1}$ contains an
        even number of $R$'s. For this configuration, $u_i=0$
        and $t_i=1$, which informs $GON(P)<GON(Q)$ and subsequently
        $P<_{GON} Q$.
        \item $P$ of length $i$, $q_{i-1}=R$ and $p_0p_1\ldots p_{i-2}$ contains an
        even number of $R$'s. Since $P$ has $i$ symbols, it means
        $u_{i-1}=1$. On the other hand, $t_{i-1}=1$ and three possible situations are posible
        \begin{enumerate}
            \item $Q$ is of length $j$ for $j>i$. Then $t_{j-1}=1$
            implies $GON(P) < GON(Q$.
            \item $Q$ is infinite-length and $q_i=L$, implying
            $t_i=1$ and $GON(P) < GON(Q)$.
            \item $Q$ is infinite-length and $q_i=R$. In this
            case there exists $j>i$ such that $q_{j}=R$.
            Otherwise, the condition $P<_{\mathcal{S}}Q$ implies
            that $P$ is of length $1$ and $Q=RLLLL\ldots$. In each of
            these situations it is satisfied $GON(P) <
            GON(Q)$.
        \end{enumerate}
    \end{enumerate}
On the other hand, let us assume $P <_{GON} Q$ and $i$ the first
index such that $u_{i}\neq t_{i}$.
    \begin{enumerate}
        \item $u_i=0$, $t_i=1$ and $p_0p_1\ldots p_{i-1}$ contains
        an odd number of $R$'s. Since $GON(P)<GON(Q)$, then $p_i=R$ and
        $q_i=L$, which further implies that $P<_{\mathcal{S}}Q$.
        \item $u_i=0$, $t_i=1$ and $p_0p_1\ldots p_{i-1}$ contains
        an even number of $R$'s. In this situation the assumption $GON(P)<GON(Q)$
        forces $p_i=L$ and $q_i=_R$, which informs that $P<_{\mathcal{S}} Q$.
        \item $P$ is of length $i$, $t_{i-1}=1$. This implies that $p_0p_1\ldots p_{i-2}$ contains
        an even number of $R$'s, $q_{i-1}=R$ and thus
        $P<_{\mathcal{S}}Q$.
        \item $u_{i-1}=0$ and $Q$ of length $i$ and $p_0p_1\ldots p_{i-1}$ contains an odd number of
        $R$'s. Therefore, $p_{i-1}=R$, $q_{i-1}=C$ and
        $P<_{\mathcal{S}} Q$.
     \end{enumerate}
    As a result, $P<_\mathcal{S}Q$ if and only if
    $P <_{GON} Q$ and the proof is complete.\qed
\end{pf}

The previous proposition and Theorem \ref{theorem:order0} lead to
the next theorem, which represents the extension and proof of
Theorem 1 in \cite{wu04}.
\begin{thm1}
    For $f\in \mathcal{F}$ and $x,y \in I$, it is satisfied
that $GON(P_f(x))\leq GON(P_f(y))$ if and only if $x\leq y$. In
other words, the GON of the symbolic sequences in $\mathcal{S}$ is
an increasing function with respect to the initial condition.
\label{theorem:orderGray}
\end{thm1}

\section{Gray codes and parametric unimodal maps}
\label{section:controlParameter} A special case of interest is the
study of unimodal maps defined in a parametric way. In this sense,
this section is focused on the analysis of the class of functions
$f_{\lambda}(x)\in \mathcal{F}$ for all $\lambda$ in $[0,1]$. Let
$F(x)\in \mathcal{F}$ and $F(x_c)=F_{\max}\leq  b$. The parametric
function $f_{\lambda}$ can be expressed as follows:
\begin{equation}
    f_{\lambda}(x)=\lambda F(x),
\end{equation}
which implies $f_{\lambda}(x_c)=\lambda \cdot F_{max}$, which is the
maximum value of $f_{\lambda}(x)$. A first consequence of this is
Theorem 3 in \cite{wu04}, which is a corollary of Theorem
\ref{theorem:orderGray}.
\begin{cor1}
For $f_\lambda(x) = \lambda F(x)$ with $F(x) \in \mathcal{F}$ and
$\lambda \in [0,1]$, it is satisfied that
$GON(P_{f_\lambda}(f_\lambda(x)))\leq
GON(P_{f_{\lambda}}(f_{\lambda}(x_c))), \forall x \in [a,b]$.
\end{cor1}
Moreover, the maximum value of $f_{\lambda}(x)$, i.e., $\lambda
F_{\max}$ depends on $\lambda$ in such a way that an increment of
the control parameter forces an increment of the maximum value. As a
consequence, the GON of the kneading sequences derived from
$x=f_{\lambda}(x_c)$ is an increasing function with respect to the
control parameter \cite[Theorem 4]{wu04}.
\begin{cor1}
For $f_\lambda(x) = \lambda F(x)$ with $F(x) \in \mathcal{F}$ and
$\lambda_1,\lambda_2 \in [0,1]$ with $\lambda_1<\lambda_2$, it is
satisfied that $GON(P_{f_{\lambda_1}}(f_{\lambda_1}(x_c)))\leq
GON(P_{f_{\lambda_2}}(f_{\lambda_2}(x_c)))$.
\end{cor1}

On the other hand, after a certain number of transient iterations,
all the values obtained from any initial condition through the
iteration of any function in $\mathcal{F}$  are inside the interval
$[x_{\min},x_{\max}]$. Therefore, once all the values derived from
the iteration of the considered function are inside
$[x_{\min},x_{\max}]$, it is verified that
$GON(P_{f_{\lambda}}(x))\geq
GON(P_{f_{\lambda}}(f_{\lambda}^{(2)}(x_c))$. This was wrongly
interpreted in \cite[Theorem 5]{wu04}, since this theorem is only
satisfied if $f_{\lambda}^{(2)}(x) \geq x_{\min}$ for any $x\in
[a,b]$. Nevertheless, the previous comments point out that this
inequality is verified only for $x\in
[f_{\lambda}^{-1}(f^{-1}_{\lambda}(x_{\min})),b]$, i.e., Theorem 5
in \cite{wu04} is not fulfilled for $x\in
[a,f_{\lambda}^{-1}(f^{-1}_{\lambda}(x_{\min}))]$. Consequently, it
is necessary to modify Theorem 5 in \cite{wu04} according to the
preceding considerations. In this sense, the next corollary rewrites
Theorem 5 in \cite{wu04} in a more accurate way and, at the same
time, extends its application domain to all the functions in
$\mathcal{F}$.
\begin{cor1}
Let $F(x)$ be a function  in $\mathcal{F}$ that leads to
$f_\lambda(x) = \lambda F(x)$ for $x \in [a,b]$ and $\lambda \in
[0,1]$. Let $x_i$ be defined as $x_i=x$ for $i=0$ and
$x_{i}=f_{\lambda}(x_{i-1})$ for $i>0, i\in \mathbb{N}$. There
exists $n_1\in \mathbb{N}$ such that $x_i$ is in
$[x_{\min},x_{\max}]$ for $i>n_1$ and it is satisfied that
$GON(P_{f_\lambda}(x_{i}))\geq
GON(P_{f_{\lambda}}(f_{\lambda}^{(2)}(x_c)), \forall x \in [a,b]$
for $i>n_1$.
\end{cor1}

Finally, the value $x_{\min}$ is given by $f^{(2)}_\lambda(x_c)=
f_{\lambda}(f_{\lambda}(x_c))=\lambda \cdot F(\lambda F_{\max})$. If
$x_{\min}$ is a monotonic function of $\lambda$, then it is possible
to extract a new corollary from Theorem \ref{theorem:orderGray}. In
\cite[Theorem 6]{wu04} it is assumed without proof that
$f_{\lambda}^{(2)}(x_c)$ is a monotonic decreasing function with
respect to $\lambda$. This assumption implies that
\begin{equation}
    \frac{\partial{x_{\min}}}{\partial{\lambda}}=F(\lambda F_{\max})+\lambda
    \cdot F_{\max}\cdot \left.\frac{\partial{F(x)}}{\partial{x}}\right|_{x=\lambda
    F_{\max}}<0.
    \label{equation:monotonicCondition_xmin}
\end{equation}
This condition is not satisfied for all the possible values
$\lambda$ and for all the functions in $\mathcal{F}$. Let us
consider the logistic map. In \cite{wu04} the dependency of
$x_{\min}$ on $\lambda$ is  studied using the logistic map. Indeed,
the logistic map is a function included in $\mathcal{F}$, which is
defined as
\begin{equation}
    f_{\lambda}(x) = \lambda \cdot 4x(1-x),
    \label{equation:logisticMap}
\end{equation}
for $\lambda \in [0,1]$ and $x\in [0,1]$. It is easy to verify that
for the logistic map the condition given by
Eq.~\eqref{equation:monotonicCondition_xmin} is fulfilled if and
only if $\lambda > 8/12$. Therefore, Theorem 6 in \cite{wu04} must
be rewritten in such a way that the discussed inaccuracy is overcome
and, simultaneously, the application domain of its variant affects
not only the logistic map but all the functions in $\mathcal{F}$.
Again, this aim is completed through a series of additional
assumptions on the scope defined in Theorem \ref{theorem:orderGray}.

\begin{cor1}
Let us suppose that $f_\lambda(x) = \lambda F(x)$ with $F(x) \in
\mathcal{F}$, $\lambda \in [0,1]$ and $x \in [a,b]$. For
$\lambda_1,\lambda_2 \in [0,1]$ with $\lambda_1<\lambda_2$ and
satisfying $\partial{f_\lambda^{(2)}(x_c)/\partial{\lambda}}<0$ for
$\lambda=\left\{\lambda_1,\lambda_2\right\}$, it is verified that
$GON(P_{f_{\lambda_1}}(f_{\lambda_1}^{(2)}(x_c)))\geq
GON(P_{f_{\lambda_2}}(f_{\lambda_2}^{(2)}(x_c)))$.
\end{cor1}

\section{Conclusions}
\label{section:conclusions} In this paper we have mathematically
proven that it is possible to read the \emph{classical} theory of
applied symbolic dynamics for unimodal maps from the point of view derived
from the concept of Gray Ordering Number. Indeed, the main results
of the present work were previously presented in other works as
theorems. Nevertheless, these theorems were not formally
demonstrated. We have provided not only the mathematical proof of these
theorems but also solved some imprecisions, which is essential to
use the concept of Gray Ordering Number in a correct and efficient
way. The main result of all this work is the possibility of
improving and expanding previous contributions based on the concept of Gray
Ordering Number. Specially relevant is the case of the estimation of
the values of the initial condition and the control parameter of
unimodal maps. The theoretical framework presented in this paper
allows to establish the limitations of the methods previously
proposed for the estimation of those values. Furthermore, this paper is
the theoretical conclusion of all the work that we have carried out on
unimodal maps both in the field of
the applied theory of symbolic dynamics
\cite{alvarez98,alvarez98a,arroyo09c}, and in the context of
chaos-based cryptography
\cite{alvarez03a,arroyo09,rhouma09,arroyo11}. 

\section*{Acknowledgments} 

This work was supported by the Spanish Government project BFU2009-08473. The work of David Arroyo was
supported by a Juan de la Cierva fellowship from the Ministerio de
Ciencia e Innovaci\'on of Spain.

\end{document}